\documentclass[hf
]{ceurart}

\usepackage{listings}
\begin{document}

\copyrightyear{2024}
\copyrightclause{Copyright for this paper by its authors.
  Use permitted under Creative Commons License Attribution 4.0
  International (CC BY 4.0).}

\conference{Joint Proceedings of the ACM IUI Workshops 2024, March 18-21, 2024, Greenville, South Carolina, USA}

\title{Human-AI Co-Creation of Worked Examples for Programming Classes}

% \tnotetext[1]{You can use this document as the template for preparing your
  % publication. We recommend using the latest version of the ceurart style.}

\author[1]{Mohammad Hassany}[%
orcid=0009-0004-8893-8454,
email=moh70@pitt.edu,
url=https://github.com/mhassany-pitt/,
]
\address[1]{University of Pittsburgh, Pittsburgh, PA, 15260}

\author[1]{Peter Brusilovsky}[%
orcid=0000-0002-1902-1464,
email=peterb@pitt.edu,
url=https://sites.pitt.edu/~peterb/,
]

\author[2]{Jiaze Ke}[%
orcid=0009-0003-3122-2298,
email=jiazek@andrew.cmu.edu,
]
\address[2]{Carnegie Mellon University, Pittsburgh, PA, 15213}

\author[1]{Kamil Akhuseyinoglu}[%
orcid=0000-0002-7761-9755,
email=kaa108@pitt.edu,
]

\author[1]{Arun Balajiee Lekshmi Narayanan}[%
orcid=0000-0002-7735-5008,
email=arl122@pitt.edu,
]

% \cortext[1]{Corresponding author.}
% \fntext[1]{These authors contributed equally.}

\begin{abstract}
Worked examples (solutions to typical programming problems presented as a source code in a certain language and are used to explain the topics from a programming class) are among the most popular types of learning content in programming classes. Most approaches and tools for presenting these examples to students are based on line-by-line explanations of the example code. However, instructors rarely have time to provide line-by-line explanations for a large number of examples typically used in a programming class. In this paper, we explore and assess a human-AI collaboration approach to authoring worked examples for Java programming. We introduce an authoring system for creating Java worked examples that generates a starting version of code explanations and presents it to the instructor to edit if necessary. We also present a study that assesses the quality of explanations created with this approach.
\end{abstract}


\maketitle
% TOCONSIDER (reviewer): "for a full paper submission I would wonder how coding examples of different complexity or different length influence the collaboration between instructor and AI. I assume that the longer the coding examples are the more instructors would use the generated examples. I wonder if longer or more frequent explanations are also prone to error."

% TOCONSIDER (reviewer): "One thing that I do wonder is if over time the system can learn from the instructor's explanations and generate better explanations. I could imagine over time that there could be fine-tuned models that generate better explanations, thus moving this from a human-AI co-creation problem to a n AI-only problem that could benefit students more. There is nothing in the paper to change here, just a thought."

\section{Introduction}

% Peter -- Motivation

Program code examples play a crucial role in learning how to program~\citep{linn1992CACM}.
Instructors use examples extensively to demonstrate the semantics of the programming language being taught and to highlight the fundamental coding patterns. Programming textbooks also pay a lot of attention to examples, with a considerable textbook space allocated to program examples and associated comments~\citep{deitel1994,kelley1995}. 
A typical worked example presents a code for solving a specific programming problem and explains the role and function of code lines or code chunks. In textbooks, these explanations are usually presented as comments in the code or as explanations on the margins. 
While informative, this approach focused on passive learning, which is known for its low efficiency. Recognizing this problem, several research teams developed learning tools that offered more interactive and engaging ways to learn from examples~\citep{brusilovsky2009problem,CODECAST2017,khandwala2018codemotion,park2018elicast,Hosseini2020}. 

The example-focused learning tools demonstrated their effectiveness in classroom studies, but their use by programming instructors is still limited due to the insufficient number of worked examples offered by these tools.
Although the authors of these tools usually provide a good set of worked examples that can be presented through their tools, many instructors prefer to use their own favorite code examples. The instructors are usually happy to broadly share the code of examples they created (usually providing it on the course web page), but they rarely have time or patience to augment examples with explanations and add their examples to an example-focused interactive system. Indeed, producing a single explained example could take 30 minutes or more, since it requires typing an explanation for each code line~\citep{brusilovsky2009problem,Hosseini2020} or creating a screencast in a specific format~\citep{CODECAST2017,park2018elicast}. 

This issue has been recognized by several research teams that have offered several ways to address the lack of content. Among the approaches explored are learner-sourcing, that is, engaging students in creating and reviewing explanations for instructor-provided code~\citep{hsiao2011role} and automatic extraction of information content from available sources, such as lecture recordings~\citep{khandwala2018codemotion}. In this paper, we present an alternative approach to address the lack of worked examples based on human-AI collaboration. With this approach, the instructor provides the code of one of their favorite examples along with the statement of the programming problem it is solving. The AI engine based on large language models (LLM) examines the code and generates explanations for each code line. The explanations could be reviewed and edited by the instructor. To support and explore this authoring approach, we created an authoring system, which radically decreases the time to create a new interactive worked example. The examples created by the system could be uploaded to an example-exploration system such as WebEx~\citep{brusilovsky2009problem} or PCEX~\citep{Hosseini2020} or exported in a reusable format. To assess the quality of the resulting examples, we performed a user study in which TAs and students compared code explanations created by experts through a traditional process with examples created by AI to contribute to human-AI collaborative process.

The remainder of the paper is structured as following. We start by reviewing related work, introduce the example authoring system that implements the proposed collaborative approach, and explain how specific design decisions were made through several rounds of internal evaluation. Next, we explain the design of our user study and review its results. We conclude with a summary of the work and plans for future research.

\section{Related Work}

\subsection {Worked Examples in Programming}
Code examples are important pedagogical tools for learning programming. Not surprisingly, considerable efforts have been devoted to the development of learning materials and tools to support students in studying code examples.
For many years, the state-of-the-art approach for presenting worked code examples in online tools was simply code text with comments \citep{linn1992CACM,davidovic2003learning,morrison2016subgoals}. More recently, this approach has been enhanced with multimedia by adding audio narrations to explain the code \citep{ericson2015analysis} or by showing video fragments of code screencasts with the instructor's narration being heard while watching code in slides or an editor window \citep{CODECAST2017,khandwala2018codemotion}. Both ways, however, support \emph{passive} learning, which is the least efficient approach from the prospect of the ICAP framework~\citep{chi2018icap}\footnote{The ICAP framework differentiates four modes of engagement, behaviorially exhibited by learners: \emph{passive, active, constructive and interactive}.}

An attempt to make learning from program construction examples \emph{active} was made in the WebEx system, which allowed students to interactively explore instructor-provided line-by-line comments for program examples via a web-based interface \citep{brusilovsky2009problem}. More recently, several projects~\citep{khandwala2018codemotion,park2018elicast,Hosseini2020} augmented examples with simple problems and other constructive activities to elevate the example study process to the \emph{interactive and constructive} levels of the ICAP framework, known as the most pedagogically efficient. 

\begin{figure}
    \centering
    \includegraphics[width=1\linewidth]{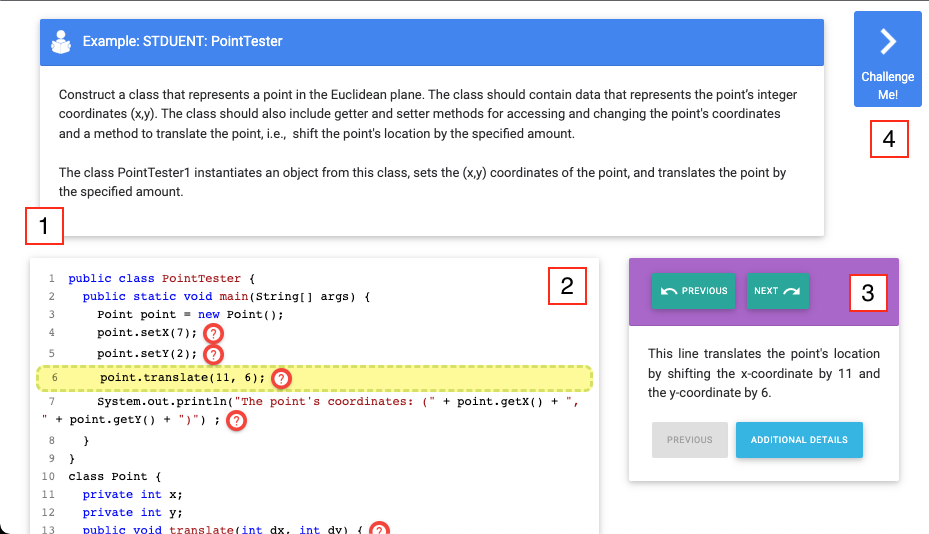}
    \caption{Studying a code example in the PCEX system: 1) title and program description, 2) program source code with lines annotated with explanations, 3) explanations for the highlighted line, 4) link to a ``challenge'' - a small problem related to the example.}
    \label{fig:pcex-example}
\end{figure}

A good example of a modern interactive tool for studying code examples is the PCEX system~\cite{Hosseini2020}. PCEX (Program Construction EXamples) was created in the context of an NSF Infrastructure project (https://cssplice.org) with a focus on broad reuse and has been used by several universities in the US and Europe in the context of Java, Python, and SQL courses.
PCEX interface (Figure~\ref{fig:pcex-example}) provides interactive access to traditionally organized worked examples, i.e., code lines augmented with instructor's explanations. Separating explanations (Figure~\ref{fig:pcex-example}-3) from the code (Figure~\ref{fig:pcex-example}-2), allows students to selectively study explanations for code lines they want. Explanations are provided on several levels of detail, so more details could be requested if the brief explanation is not sufficient (Figure~\ref{fig:pcex-example}-3). %A link to a ``challenge'' Figure~\ref{fig:pcex-example}-4) allows students to check their understanding by solving a simple problem. 

Since line-by-line multi-level example explanations offered by PCEX is currently the most detailed approach for explaining worked examples, we selected the code example structure implemented by PCEX as the target model for our authoring tool presented in this paper. The tool produces code augmented with line-by-line explanations on several levels of detail. The resulting example could be directly uploaded to PCEX or exported in a system-independent format to be uploaded to other example exploration systems like WebEx~\citep{brusilovsky2009problem}.

\subsection{Use of LLMs for Code Explanations}
% Creating educational content is challenging and time-consuming. 
Several research teams explored the use of LLM for code explanations using GPT-3~\cite{10.1145/3544548.3581388,10.1145/3545945.3569785,leinonen2023comparing}, GPT-3.5~\cite{10.1145/3545945.3569785,li2023explaining,10.1007/978-3-031-36336-8_50}, GPT-4~\cite{li2023explaining}, OpenAI Codex \cite{10.1145/3501385.3543957,tian2023chatgpt,10.1145/3545945.3569785}, and GitHub Copilot \cite{10.1007/978-3-031-36336-8_50}. 
LLMs were used to generate explanations at different levels of abstraction (line-by-line, step-by-step, and high-level summary). Sarsa et al.~\cite{10.1145/3501385.3543957} observed that ChatGPT can generate better explanations at low-level (lines). %Li et al.~\cite{li2023explaining} used the result of specific-to-general generated explanations as one of the inputs to their LLM solver, trying to solve competitive-level programming problems more efficiently. %Since ChatGPT knowledge is limited up to 2021, Tian et al.~\cite{tian2023chatgpt} highlighted that it will not perform well with unseen programming problems. To address this issue, Chen et al.~\cite{10.1007/978-3-031-36336-8_50} enriched the prompt given to ChatGPT by including relevant source files in the current project directory. 
%A novel research~\cite{10.1145/3544548.3581388} tried to understand how non-expert approach LLMs. They have identified common mistakes and provided advice for tool designers.
Explanations and summaries generated by these LLMs were mostly evaluated by authors~\cite{10.1145/3501385.3543957}, students~\cite{10.1145/3545945.3569785,leinonen2023comparing}, and tool users~\cite{10.1007/978-3-031-36336-8_50}. Sarsa et al.~\cite{10.1145/3501385.3543957} reported a high correct ratio for generated explanations with minor mistakes that can be resolved by the instructor or teaching assistant. Students rated LLM-generated explanations as being useful, easier, and more accurate than learner--sourced explanations~\cite{leinonen2023comparing}. %However, Tian et al.~\cite{tian2023chatgpt} observed that ChatGPT doesn't perform well with incorrect programs, even when only minor changes are required to correct the program. 

% TOIMPROVE (reviewer): "I would suggest that the authors consider rewriting the second paragraph in section 2.2.   It represented a sudden unmotivated shift in level of detail, and was not well introduced, and ultimately didn't seem to be making a coherent point.  You want to explain that in order to get the llm to generate the desired code explanations, the llm must be prompted with textual instructions describing what it is to do and what form those explanations should take, perhaps providing some examples.  The specific form of the prompt is critical to eliciting the desired behavior from the llm.   I believe that was the message that the authors were trying to get across there."

Since prompts directly influence the LLM’s performance, several studies focused on exploring different prompting strategies~\cite{Zhou2022LeasttoMostPE,white2023prompt}.  Tian et al~\cite{tian2023chatgpt} reported that a verbose prompt will limit the LLM's ability to utilize its knowledge~\cite{tian2023chatgpt}. Iterative prompts are proven to perform well~\cite{10.1145/3544548.3581388}. Zamfirescu-Pereira et al.~\cite{10.1145/3544548.3581388} observed that non-experts have misconceptions about LLMs and struggle to come up with a well-formed prompt. Researchers believe that LLMs can be beneficial in environments where humans and AI can work together, where the human can perform the expert evaluation and tune the responses generated by the AI while the AI performs the time-consuming manual tasks \citep{white2023prompt}.

\section{The Feasibility Studies}
% Arun
%Arun, your task is to expand the section on “feasibility studies” that I just added before the interface description. We need to cite two studies - the one with Memphis from AAAI workshop, which shows that LLM explanations are harder to read and further away from students and the other that is now SAC poster (and arXiv paper) where we demonstrated that LLM explanations assessed quite positively by humans. The conclusion from these two studies, which justify our interface design, are already written for me, but now we need some data from the studies, basically, a paragraph of two for each, with a table. For AAAI study where the results are concisely presented in two tables, you could simply reuse the tables, removing all lines focused on specific prompts and leaving only average ChatGPT, students, and experts data. For SAC posters, you should try to condense the results attempting to present the “gist” in one or two tables. The current way we use to present these results is too verbose, see how you could present it better.

To assess the feasibility of Human-AI co-creation of worked examples, we performed three rounds of preliminary studies. The purpose of these studies was to develop an approach for producing LLM code explanations of reasonable quality, compare the explanations produced by LLMs with the explanations produced by humans, and assess whether the LLM explanations are considered satisfactory by instructors and students. 

In the first study~\cite{hassany2023arhiv} guided by earlier work on LLM code explanations reviewed above, we explored a range of prompts and performed an evaluation of the quality of explanations generated by the prompts to select the best-performing prompt for the next rounds of our work. 

In the second study~\cite{lekshmi2024explaining}, we used a dataset of explanations produced by two experts and 60 students for the same four Java code examples with 33 explainable lines to compare ChatGPT explanations with explanations produced by experts and students using several formal metrics. To make this comparison, we generated ChatGPT explanations using our selected prompt for the 33 explainable lines four times, using temperature 0 once and temperature 1 three times. To calculate all comparison metrics, we merged all line explanations generated by each source (i.e, each expert, each student, and each round of ChatGPT generation) into a single source document. As the data shows (Table~\ref{tab:lexcial_metric_detail}), the explanations produced by ChatGPT have comparable length (measured by the number of tokens) and lexical density with the explanations produced by experts, while the explanations produced by students were more than twice as short and more lexically dense than the explanations produced by the other two sources. Surprisingly (given the length difference) the readability of explanations produced by experts is very similar to the readability of student explanations, while ChatGPT explanations are much less readable. Expert explanations are also much more similar than ChatGPT explanations to the explanations produced by students (Table~\ref{tab:similarity_metric_table}). This data could be partially explained by the considerably larger vocabulary used by ChatGPT even in comparison to experts.  

% -- TOIMPROVE (reviewers): "while I’m not familiar with all the metrics provided in Table.1 and Table. 2"

%Arun
\begin{table}[h!]
    
    \begin{tabular}{c c c c c c c c }
    \hline
         Source &  N & Vocabulary & Lexical Density  & \# of Tokens & GF & FRE & FK\\
         \hline
            Experts & 2 & 209.0  & 0.48 & 690.0 & 8.46 & 78.45 & 6.18\\
            ChatGPT* & 4 & 238.0  & 0.49 & 769.5 & 11.09 & 69.64 & 7.83	\\
            Students & 60 & 116.5 & 0.54 & 249.5 & 8.02 & 80.48 & 5.62 \\
\hline
\end{tabular}
    \caption{Median lexical and readability metrics for different sources of explanations (FRE = Flesch-Reading Ease, FK = Flesch-Kincaid, GF = Gunning Fog). *refers to the prompt selected in the first study.} 
    \label{tab:lexcial_metric_detail}
\end{table}

\begin{table}[h!]
    \centering
    \begin{tabular}{ c c c c c c}
    \hline
        Reference & Source & chrF & METEOR & USE & BERTScore \\
    \hline
        Expert & Student  & 0.33 & 0.144 & 0.33 & 0.63 \\
        ChatGPT & Student & 0.18 & 0.151 & 0.255 & 0.458 \\
                Expert & ChatGPT  & 0.32 & 0.28 & 0.48 & 0.712 \\
    \hline
    \end{tabular}
    \caption{Assessing lexical and semantic alignment (larger is better) between sources of explanations. }
    \label{tab:similarity_metric_table}
\end{table}

%%% ============== first study ends 

In the third study~\cite{hassany2023arhiv}, we conducted a comparative evaluation of explanations produced by experts and ChatGPT from the point of view of human users. We used two types of human users: authors (instructors and TAs) who are expected to use ChatGPT-generated explanations as the starting point in the co-creation process, and students who are the target users of the co-created product.  Explanations were compared in pairs, each explanation in a pair has to be judged by completeness, and the best explanation in the pair has to be selected. A pair included an expert and a ChatGPT explanation, and the judges were not aware of which source produced each explanation. The study results indicated strong preferences for ChatGPT in both groups of judges (Table~\ref{tab:chatgpt-expert_completeness-whichisbetter-avgstdev}). In general, ChatGPT explanations were rated as more complete and judged to be better in the majority of cases. However, it was not a clear win. In a substantial number of cases (15.05\% for students and 27.41\% for authors), expert explanations were selected as the best option in a pair.

%Arun

\begin{table}[t]
\begin{tabular}{@{}llcccc@{}}
\toprule
\ \ Source & Judged by & Not complete & Complete & Very complete & ``This source is better''\\ \midrule

\ \ ChatGPT & Students & 0.00\% & 13.33\% & 86.67\% & 51.11\% \\
\ \ ChatGPT & Authors & 1.48\% & 32.59\% & 65.93\% & 58.15\%  \\
\hline

\ \ Experts & Students & 2.22\% & 55.56\% & 42.22\% & 16.05\%* \\
\ \ Experts & Authors & 14.07\% & 57.78\% & 28.15\% & 27.41\%* \\
\bottomrule
\end{tabular}
% \vspace{0.2cm}
\caption{Assessment of explanations generated by ChatGPT and experts by students and authors. For convenience, we do not count the cases in which the explanations in a pair were judged equally good.}
\label{tab:chatgpt-expert_completeness-whichisbetter-avgstdev}
\end{table}

% try to merge tables 4 and 5 - just tell what students and authors tells about (1) completeness - 3 columns and (3) who is better - again 3 columns. It should fit

% \begin{table}[t]
% \begin{tabular}{@{}lcc@{}}
% \toprule
% \cmidrule{2-3}
% Rating & Students & Authors \\
% \hline
% Both are the same = 0 & 32.84\% & 14.44\% \\
% Expert is better = 1 & 16.05\% & 27.41\%  \\
% ChatGPT is better = 2 & 51.11\% & 58.15\% \\
% \bottomrule
% \end{tabular}
% % \vspace{0.2cm}
% \caption{Percentage of Ratings for the different items on the scale for ``Which explanation is better?''}
% \label{tab:chatgpt-expert_whichisbetter}
% \end{table}

% \begin{table}[]
% \begin{tabular}{@{}ccc@{}}
% \toprule
% \multicolumn{1}{l}{} & Students & Authors \\ \midrule
% ChatGPT* &  1.644 (0.258) & 1.778 (0.163) \\
% Expert* &  1.141 (0.465) & 1.296 (0.408) \\
% Which is better? &  1.437 (0.373) & 1.284 (0.427) \\ \bottomrule
% \end{tabular}
% % \vspace{0.2cm}
% \caption{Average (Stdev) Ratings - *Completeness}
% \label{tab:avgstdev}
% \end{table}

%%% === second study ends

Taking the results of these two studies together, we could conclude that producing explanations for code examples is a promising application area for Human-AI co-creation. On the one hand, the LLM-generated explanations are lagging behind expert explanations in several aspects. ChatGPT explanations have higher reading difficulty than expert explanations, and they are further away from the students' own explanations, as measured by most similarity metrics. The vocabulary data hints that ChatGPT tends to use terms, which might not be easy for the students to understand, while experts have experience in phrasing their explanations closer to the students' active vocabulary. On the other hand, the explanations produced by ChatGPT were generally rated higher than the expert explanations by both instructors and students. These data hint that presenting ChatGPT explanations directly to students might not be a perfect solution, but they can serve as an excellent starting point for instructors in shaping their own explanations. Following that, we decided to structure the Human-AI collaboration in creating working examples as follows. Instructors have the ultimate control over producing explanations. Depending on the context (such as example complexity), they can either choose to explain example lines themselves or request AI (LLM) help in producing explanations for specific lines. In the latter case, LLM generates the initial line explanations leaving it to the instructor to accept or reject it and, if accepted, to further edit the explanation text to satisfaction. The Human-AI co-creation interface presented in the next section is based on this model of collaboration.
%\section{Programming Construction Worked Examples}

% Peter / Kamil

\section{The Human-AI Co-Creation Interface Design}

On the basis of our feasibility studies, we developed a Worked Example Authoring Tool (WEAT). WEAT enables instructors to create worked code examples for PCEX system,~\cite{Hosseini2020} through the human-AI co-creation interface. In this co-creation process, the main task of a human author is to provide the code of the example and the statement of the problem that the code solves. The main task of ChatGPT is to generate the bulk of code line explanations on several levels of detail. As an option, a human author could edit and refine the text produced by ChatGPT to adapt it to the class goals and target students. As in any productive collaboration, each side does what it is best suited to do, leaving the rest to the partner. 

\begin{figure}[h]
    \centering
    \includegraphics[width=1\linewidth]{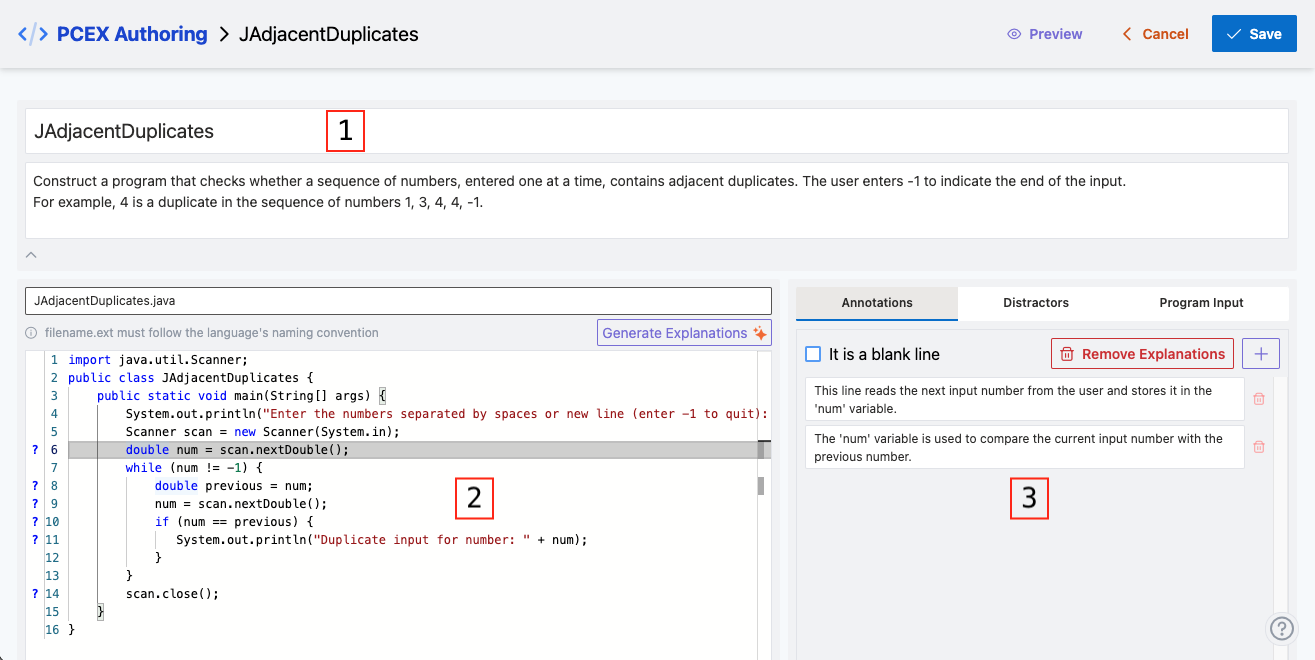}
    \caption{WEAT Authoring, 1) program title and description, 2) program source code (lines with explanations are marked with a blue question mark next to the line number), 3) explanations for the selected line (the line with gray background - line 6 in the screenshot).}
    \label{fig:pcex-authoring}
\end{figure}

In the main part of the WEAT interface, the problem  (Figure \ref{fig:pcex-authoring}-1) and the code (Figure \ref{fig:pcex-authoring}-2) have to be provided by the instructor, while the explanations for each line (Figure \ref{fig:pcex-authoring}-3) can be created by the instructor or generated by ChatGPT. The generated explanations could be further edited by the instructor.
%!, who could also turn a regular example into a challenge by marking some lines as blank (within a challenge, these lines can be replaced with distractors, one-line code snippets, by the student). 
While we expect that co-creation of code explanations will be the preferred way to use WEAT, the system supports the whole range of options from using AI explanations without human editing to creating the whole example from scratch, without the help of AI. Authors who want to start by creating explanations themselves could simply select a code line to explain (Figure \ref{fig:pcex-authoring}-2) and add one or more explanation fragments to this line (Figure \ref{fig:pcex-authoring}-3). The order of the fragments is important: the first fragment is displayed in PCEX when the line is clicked, while the remaining fragments can be accessed by clicking the ``Additional Details'' button (Figure \ref{fig:pcex-example}-3).
%Although our authoring tool addresses most of the authoring challenges, providing explanations is a time-consuming task. In the context of our tool, LLMs like ChatGPT can be used to generate line-by-line code explanations. 

%\subsection{Mechanism}

% Creating a worked example with the help of ChatGPT consists of two steps: 1) identify lines that need to be explained, and 2) generate explanations for these lines. Program description can provide a rich context for these two steps.

\begin{figure}
    \centering
    \includegraphics[width=1\linewidth]{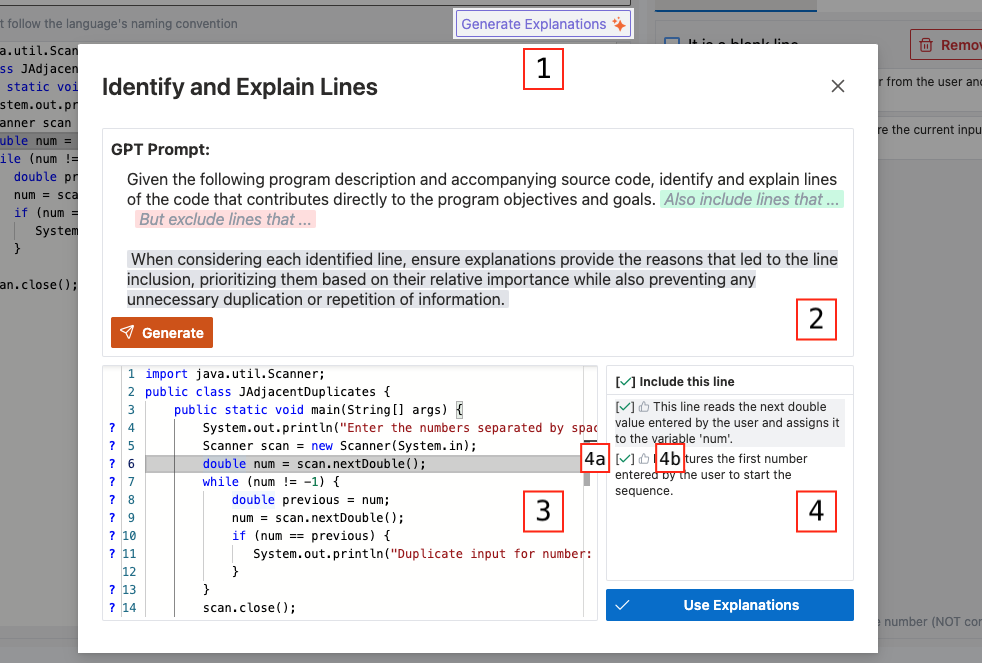}
    \caption{Human-AI Collaborative Worked Example Authoring, 1) ``Generate Explanations'' button, 2) default prompt (author can tune the prompt - optional), 3) program source preview, 4) generated explanations for the selected line.}
    \label{fig:human-ai_pcex-authoring}
\end{figure}

To generate ChatGPT explanations for the provided example code and problem description, the author has to click the ``Generate Explanations'' button to open the ChatGPT dialog (Figure \ref{fig:human-ai_pcex-authoring}). In this dialog, the explanations could be generated by clicking ``Generate'' button and added to the example by clicking ``Use Explanations'' button. Experienced authors have the opportunity to tune the default prompt before generating explanations and review the generated explanations before using them. Reviewing the generated explanations can be done line by line: selecting one of the explained lines (marked by ``?") in the code box (Figure \ref{fig:human-ai_pcex-authoring}-3) will display all generated explanations for this line in the explanation box (Figure \ref{fig:human-ai_pcex-authoring}-4). The explanation could be accepted or rejected by clicking the checkbox next to the ``Include this line'' prompt.

To support the review at the finer grain level, WEAT divides the explanations into fragments that can be independently accepted or rejected by clicking the small green check mark icon next to the fragment (Figure \ref{fig:human-ai_pcex-authoring}-4a). The author can also click on the small gray thumb-up icon (Figure \ref{fig:human-ai_pcex-authoring}-4b) to provide positive feedback on the explanation fragment. Once the ``Use Explanations'' button is clicked, all accepted explanation fragments are added to the corresponding example lines and can be further edited in the main interface (Figure \ref{fig:pcex-authoring}).
%The WEAT provides the human author several opportunities to control the outcome of the explanation generation process: 1) the author can tune the prompt to their needs, 2) the author can decide whether to include or exclude a generated explanation and 3) the author can edit or remove the explanation after it is edited.

\section{Evaluation}

% should we add a link to our authoring tool (promotion)?!

To assess how well WEAT supports co-creation of worked examples, we engaged five instructors  (A1-A5) teaching Java of Python classes and asked them to create one or more worked examples for PCEX from real examples they use in their classes. To explain the tool to the instructor, we provided a video tutorial and integrated textual help into WEAT. Their interactions and usage of the tool were recorded through logs and used for the analysis presented below. 

\begin{figure}[h]
    \centering
    \includegraphics[width=1\linewidth]{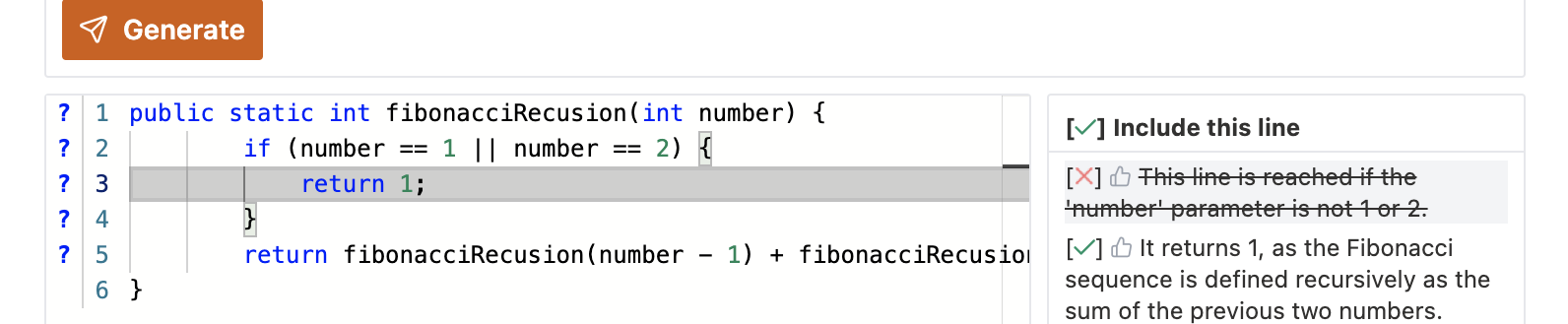}
    \caption{An incorrect explanation fragment generated by ChatGPT and excluded by the author (line 3).}
    \label{fig:a-human-ai-collaboration-example}
\end{figure}

The instructors used the tool to create 12 examples in total (Table \ref{chatgpt-used-sessions_report}). The ChatGPT dialog was used 21 times, and in 13 cases (A1=6, A2=2, A3=3, A4=1, and A5=1), instructors added generated explanations to the example by clicking the ``Use Explanations'' button. As discovered from an interview with instructors, in several cases they closed and reopened the ChatGPT dialog to access the main interface blocked by the dialog. Analyzing the interaction logs, we observed this has been done at least 5 times (3 times with the close-reopen interval of 5 seconds and 2 with 12 seconds interval) leaving only 16 cases where explanations had a chance to be examined. 
In total, 269 explanation fragments were generated for 119 lines of code with an average of 2.26 fragments per line. In 13 cases where ChatGPT explanations were added to the example by instructors, ChatGPT generated 237 explanations for 99 lines of code (Table \ref{chatgpt-used-sessions_report}). We found no cases in which the entire set of explanations generated for the line was excluded by the instructors in its entirety, and among the 237 generated fragments, only 24 (10.12\%) 237 were excluded. The interview revealed that in some cases the generated fragments were rejected not because they were unsatisfactory, but because they were incorrect (Figure \ref{fig:a-human-ai-collaboration-example}). On the other hand, instructors liked 15 (6.32\%) explanations. 

\begin{table}[h]
\centering
\begin{tabular}{l c c c c c c}
\hline
 & A1 & A2 & A3 & A4 & A5 & Total \\
\hline
Examples Created & 6 & 2 & 2 & 1 & 1 & 12 \\
Generated Explanations& 126 & 32 & 44 & 8 & 27 & 237 \\
Lines of Code being Explained by ChatGPT& 55 & 12 & 21 & 2 & 9 & 99 \\
% Excluded Lines of Code& 0 & 0 & 0 & 0 & 0 & 0 \\
Explanations Excluded & 18 & 2 & 4 & 0 & 0 & 24 \\
Explanations Liked & 6 & 0 & 9 & 0 & 0 & 15 \\
Explanations Edited & 29 & 0 & 11 & 0 & 26 & 66 \\
Explanations Removed & 8 & 0 & 15 & 0 & 0 & 23 \\
\hline
\end{tabular}
\caption{Analysis of ChatGPT used explanations: Total count of generated, excluded, liked, and explained lines of code across 13 instances where the instructor added explanations to examples.}
\label{chatgpt-used-sessions_report}
\end{table}

After adding explanations to the example, instructors still didn't remove the explanations for any line entirely, but removed 23 (9.7\%) ChatGPT generated explanation fragments. Instructor A5 reported that he removed several fragments when merging two or more explanation fragments. Since the tool did not provide support for merging fragments, it did so by copying the explanation from one fragment to the end of the other fragment and removing the obsolete fragment. In only 10 cases, instructors attempted to create new explanations from scratch, but in the end these explanations were removed. In other words, all remaining explanation fragments were originally generated by ChatGPT with some of them being edited later by the instructors. Apparently, the instructors preferred to edit the explanation fragments rather than create them from scratch. In total, the instructors edited 66 (27.84\%) of ChatGPT generated explanation fragments, on average 1.4 times (stdev=0.55). Feedback from instructors indicated that most of their edits involved summarizing, adding missing details, or removing unnecessary parts. Table \ref{chatgpt-used-sessions_report} shows that almost half of the generated fragments were used without being touched, saving a noticeable amount of instructor time. 

\begin{table}[h]
\centering
\begin{tabular}{l c c c c c c c}
    & A1 & A2 & A3 & A4 & A5 & Total \\
\hline
% Number of Instructor-added Explanations & 5 & 0 & 5 & 0 & 0 & 10 \\
ChatGPT Edited Explanations & 29 & 0 & 11 & 0 & 26 & 66 \\
ChatGPT Explanation Edits & 42 & 0 & 12 & 0 & 39 & 93 \\
\hline
Average Levenshtein Ratio across all &  &  &  &  &  & Average \\
\hspace{0.25cm}Final and Original ChatGPT Explanations & 0.435 & 1 & 0.833 & 1 & 0.412 & 0.736\\
\hline
\end{tabular}
\caption{ChatGPT-generated explanations edits: Number of edits made by instructors to ChatGPT-generated explanations, along with a measure of similarity between the original and final edited version.}
\label{post-chatgpt-session_edits-report}
\end{table}

The average Levenshtein edit ratio for ChatGPT-generated explanations (edited and unedited) is 0.73 (Table \ref{post-chatgpt-session_edits-report}), indicating a high acceptance rate for generated explanations. This indicator, however, is somewhat misleading since a portion of ChatGPT-generated explanations were edited because the first version of the tool evaluated in the study didn't provide direct support for reordering and merging the explanations, resulting in copy-pasting the explanations (as reported by A5 for whom the ratio dropped to 0.412). The table also points out that WEAT was able to support different editing approaches pursued by instructors. Some instructors spent more time reviewing the generated explanations before adding them to the example (A1), some prefer adding them to the examples and then evaluating and editing them (A3), while some used the generated explanations without changes. 

\section{Conclusion}

In this paper, we introduce a worked code example authoring tool WEAT that supports human-AI co-creation in the process of developing such examples. WEAT supports human authors by using ChatGPT for the generation of line-by-line code explanations and by providing an interface to integrate this functionality into a balanced authoring process. To the best of our knowledge, this is the first attempt to develop an authoring tool that produces worked examples through human-AI collaboration.

To develop WEAT, we performed several rounds of feasibility studies. These studies supported the need for a human-AI co-creation in authoring worked examples. As the studies showed, in the majority of cases, the explanations generated by ChatGPT with a carefully tuned prompt were positively evaluated by authors and students. However, in a good fraction of cases they were inferior to the explanations provided by experts. The study also revealed that on average experts can create explanations that are more easily readable and closer to the explanations generated by the students themselves. With this data, we hypothesized that human-AI co-creation could offer the ``best of both worlds'' solution where good explanations could be simply accepted by authors, while inferior or hard-to-understand explanations could be improved.

An evaluation of WEAT system with five course instructors supported these expectations and provided strong evidence in favor of co-creation. As the log analysis demonstrated, in many cases, instructors choose to accept generated explanations without changes, which should have decreased the time and effort required for example creation. Yet in other cases, the instructor rejected or edited the generated explanation to achieve the desired quality. In some cases, explanations were rejected by being simply incorrect, which stresses the importance of human presence in the authoring process. The interview with authors revealed several cases where authors acted inefficiently due to specific interface issues, such as blocking the main edit window by the generation dialog or the lack of tools to move or merge fragments. Now we are using these observations to develop an improved version of WEAT.

As the first step towards this important goal, our work has limitations. Most importantly, the scale of our evaluation is relatively small. Since we targeted real instructors as users in our evaluation process, we were able to recruit only five qualified subjects. Additionally, since the study was done at the beginning of the semester when instructors were busy setting up their classes, they created only 12 examples using this tool. To obtain more reliable data, we plan a larger-scale semester-long study by engaging instructors to create a variety of worked examples of varying difficulty and use them in their classes. Such a study will also enable us to assess the quality of explanations produced through human-AI collaboration and their value for students in introductory programming classes.

\bibliography{references}

\end{document}